\documentclass[manuscript]{aastex} 

\usepackage{graphicx,epsfig,xcolor,subfigure,multicol}
\usepackage{graphicx,epsfig}
\usepackage{natbib}                              
\usepackage{rotating}
\usepackage{txfonts}
\usepackage{longtable,lscape}
\usepackage{rotating}

\shorttitle{Rotation of the Optical Polarization Angle Associated with}
\shortauthors{Sorcia et al.}

\begin{document}

\title{ROTATION OF THE OPTICAL POLARIZATION ANGLE ASSOCIATED WITH THE 2008 $\gamma$-RAY FLARE OF BLAZAR W COMAE}

\author{Marco Sorcia\altaffilmark{1}, Erika
  Ben\'itez\altaffilmark{1}, David Hiriart\altaffilmark{2},
  Jos\'e M. L\'opez\altaffilmark{2}, Jos\'e I.
  Cabrera\altaffilmark{1}, and Ra\'ul M\'ujica\altaffilmark{3}}


\altaffiltext{1}{Instituto de Astronom\'ia, Universidad Nacional Aut\'onoma
  de M\'exico, Apdo. 70-264, Mexico D.F., 04510, Mexico}

\altaffiltext{2}{\noindent Instituto de Astronom\'ia, Universidad Nacional 
Aut\'onoma de M\'exico, Apdo. 810, Ensenada, B.C., 22800, Mexico}
\altaffiltext{3}{Instituto Nacional de Astrof\'isica, \'Optica y
  Electr\'onica, Apdo. Postal 51 y 216, 72000 Tonantzintla, Puebla, Mexico}


\begin{abstract}

  An $R$--band photopolarimetric variability analysis of the TeV bright blazar
  W Comae, between 2008 February 28 and 2013 May 17, is presented. 
  The source showed a gradual tendency to decrease its mean flux level with a total change
  of ~3~mJy.  A maximum and minimum brightness states
  in the $R$-band of 14.25$\pm$0.04 and 16.52$\pm$0.1~mag respectively were observed,
  corresponding to a maximum variation of $\Delta$F = 5.40 mJy.
  We estimated a minimum variability timescale of $\Delta$t=3.3 days.  A
  maximum polarization degree $P$=33.8\%$\pm$1.6\%, with a maximum variation
  of  $\Delta$P = 33.2\%, was found.  One of our main results is the detection of a large rotation 
  of the polarization angle from 78\degr to 315\degr ($\Delta\theta\sim$237\degr) that coincides in time with the $\gamma$-ray flare
  observed in 2008 June.  This result indicates that both optical and
  $\gamma$-ray emission regions could be co-spatial.  During this flare, a
  correlation between the $R$-band flux and polarization degree was found with
  a correlation coefficient of $r_{F-p}=0.93\pm$0.11. 
  From the Stokes parameters we infer the existence of two optically thin
  synchrotron components that contribute to the polarized flux.  One of them
  is stable with a constant polarization degree of 11\%. Assuming a shock-in jet
  model during the 2008 flare, we estimated a maximum Doppler factor
  $\delta_D\sim 27$ and a minimum of $\delta_D\sim 16$; a minimum viewing
  angle of the jet $\sim$2\degr.0; and a magnetic field $B \sim$ 0.12~G.

\end{abstract}

\keywords{(galaxies:) BL Lacertae objects: individual (ON231, W~Comae) ---
  galaxies: jets --- galaxies: photometry --- polarization}

\section{INTRODUCTION}

The study of the Blazar phenomenon has been one of the major topics of study
on the Active Galactic Nuclei (AGN) family because of their extreme properties.
They show strong flux variability, superluminal motion, and a non-thermal
continuum extending from radio to TeV $\gamma$--ray regions
\citep[e.g.,][]{2005AJ....130.1418J,2010ApJ...716...30A,2010ApJ...710L.126M,2013hsa7.conf..152A}.
Blazars are radio--loud AGN and consist of BL Lacertae objects and
Flat Spectrum Radio Sources
\citep[FSRQ;][]{1980ARA&A..18..321A,1997MNRAS.289..136F,2010ApJS..189....1A,2012MmSAI..83..138C}.
These properties are explained through the idea that blazars are objects with
a very small viewing angle, i.e. the emission produced by the relativistic jet
is aligned very close to the observer's line of sight
\citep[e.g.,][]{1978bllo.conf..328B,2009A&A...494..527H}. In recent years, it
has been well established that the non-thermal continuum emission in Blazars
shows two broad low-frequency and high-frequency components in their Spectral
Energy Distribution (SED).  In the case of the BL Lac objects, this empirical
property conforms the base for classifying them accordingly to the location
of the first peak, known as the synchrotron peak, in the SED
\citep{1995ApJ...444..567P,2006A&A...445..441N,2012MmSAI..83..138C}. Commonly,
low-frequency-peaked BL Lacs (LBL) have their synchrotron peak, $\nu_{\rm
  syn}^{\rm peak} < 10^{14}$ Hz; intermediate frequency-peaked BL Lacs (IBL)
in the range 10$^{14} <\nu_{\rm syn}^{\rm peak} <10^{15}$ Hz; and
high-frequency-peaked BL Lacs (HBL) have $\nu_{\rm syn}^{\rm peak} > 10^{15}$
Hz.

The blazar W Comae at z=0.102 (also known as 1219+285 or ON 231) was
discovered as a radio source by \citet {1971Natur.231..515B}.  VLBI
observations of W Comae revealed a complex jet that extends toward the east at
$\theta \sim$100$^\circ$ \citep {1992ApJ...388...40G, 1994ApJ...435..140G}.
Also, it was found that the jet shows superluminal components with strong
polarization. The polarized emission components are found to be both aligned
with and transverse to the local jet direction in different jet components
\citep {1996MNRAS.283..759G}.

The optical historical light-curve of W Comae shows variations at all scales,
from days and weeks, to months and years \citep[see
e.g.][]{1995A&A...295....1L,2000A&A...356L..21B,1998A&AS..130..109T,1999A&A...342L..49M}.
Also, it has shown rapid variations on scales of hours
\citep{2002ARep...46..609B}.  \citet{1998A&AS..130..109T} observed the highest
brightness value ever observed for W Comae since 1940, reaching a 
maximum of B = 14.2 mag in 1997 January.  Later, \citet{1999A&A...342L..49M}
reported a very strong flare of W Comae when the object reached a
historical maximum of $R\sim$12.2~mag in 1998 April 23.  Optical polarization
of W Comae was also reported in \citet{1999A&A...342L..49M}. Their multi-band
optical observations were done just before and during its brightest phase
(1998 April 17--25). During the brightest state, the polarization was low in
the $UBV$ filters ($\sim$ 2\% to 4\%) with less than 0.4\% in the $R_c$ and
$I_c$ filters.

The $\gamma$--ray emission of W Comae has been detected by the Energetic Gamma
Ray Experiment Telescope ({\it EGRET}) on board of the Compton Gamma Ray
Observatory {(\it CGRO}) in the 100~MeV to 10~GeV band \citep
{1999ApJS..123...79H}. {\it BeppoSAX} data analysis of W Comae given by
\citet{2000A&A...354..431T} demonstrates that this source is an IBL source.
This blazar was considered as a very interesting target for the very high
energy (VHE) observatories due to the possibility of being a $\gamma$-ray
source that could be detected by Cerenkov telescopes such as the High Energy
Stereoscopic System ({\it HESS}), the {\it MAGIC} telescopes, and the Very
Energetic Radiation Imaging Telescope Array System ({\it VERITAS}) \citep[see,
e.g.][]{2002ApJ...581..143B}.  This prediction was confirmed later, when W
Comae was discovered to be a $\gamma$--ray emitter at VHE by {\it VERITAS} in
2008 March 15 \citep[see][]{2008ApJ...684L..73A}.  Thus, W Comae is the first
IBL detected at VHE.  A subsequent multiwavelength campaign on this object was
coordinated during a major $\gamma$-ray flare in 2008 June
\citep{2009ApJ...707..612A}. A very high $\gamma$-ray signal was detected by
{\it VERITAS} in 2008 June 8 that was brighter, by a factor of three, than the
previous emission detected in 2008 March.

In this paper we report the results of the photopolarimetric monitoring of the
TeV--blazar W Comae carried out from 2008 February to 2013 May. Our main goal
is to establish the long--term optical variability properties of the polarized
emission in the $R$--band. The variability of the Stokes parameters obtained
from our observations is analyzed in terms of a two--component model.
Estimations of some of the physical parameters that are known to be associated with
the kinematics of the relativistic jet are obtained. One of our main results
is the detection of a large rotation of the electric vector position angle
(EVPA) that coincides with the time of occurrence of the major flare observed in
$\gamma$-rays in 2008 June 8.

This paper is organized as follows: section~\ref{Obse} presents a description
of our observations and our main observational results. Polarimetric
properties are analyzed in section~\ref{Polana}. Results are discussed in
section~\ref{Dis}. Finally, in section ~\ref{Conc} we show our conclusions.
Throughout this paper we use a standard cosmology with $H_{0}$\,=\,71km
s$^{-1}$ Mpc$^{-1}$ , $\Omega_m$ = 0.27, and $\Omega_{\Lambda}$=0.73.

\section{OBSERVATIONS AND RESULTS}
\label{Obse}

The observations were carried out with the 0.84 m f/15 Ritchey-Chretien
telescope at the Observatorio Astron\'omico Nacional of San Pedro M\'artir
(OAN-SPM) in Baja California, Mexico and the instrument POLIMA.  The
differential $R$-band magnitudes of W Comae were calculated using the standard
star~A distant about $\sim$1.2~arc-minutes to the South-East from the studied
object. The magnitude of the comparison star~A in the $R$--band is
(11.72$\pm$0.04) mag \citep{1996A&AS..116..403F}.  Because of the narrow
field of view of the instrument, $\sim$ 4~arc-minutes, this was the only
standard star available for calibration with a reasonable flux level. The
exposure time was 80~s per image for W Comae. Polarimetric calibrations were
made using the polarized standard stars ViCyg12 and HD155197, and the
unpolarized standard stars GD319 and BD+332642 \citep{1992AJ....104.1563S}.
$R$-band magnitudes were corrected for the host galaxy contribution, $m_{R(\rm
  host)}$=16.60, fitting a de Vaucouleaurs profile
\citep[see][]{2003A&A...400...95N}. Then, the magnitudes were converted into
apparent fluxes using the expression: $F_{\rm obs}=K_0 \times10^{-0.4m_R}$,
with $K_0= 3.08\times 10^6$ mJy \citep{2007A&A...475..199N}, for an effective
wavelength of $\lambda = 640 \,$nm.

The ambiguity of 180\degr  in the polarization angle was corrected in such
  a way that the differences observed between the polarization angle of
  temporal adjacent data should be less than 90\degr. We
  defined this difference as:
\begin{equation}
 |\Delta\theta_{n}|=|\theta_{n+1}-\theta_{n}| - \sqrt{\sigma(\theta_{n+1})^2+\sigma(\theta_{n})^2}, 
\end{equation}
 where $\theta_{n+1}$ and $\theta_{n}$ are the $n+1$ and n-th polarization
angles and $\sigma(\theta_{n+1})$ and $\sigma(\theta_{n})$ their errors. If
$|\Delta\theta_{n}|\leq90^{\circ}$, no correction is needed. If
$\Delta\theta_{n}\,<\,-90^{\circ}$, we add $180^{\circ}$ to $\theta_{n+1}$. If
$\Delta\theta_{n}\,>\,90^{\circ}$, we add $-180^{\circ}$ to $\theta_{n+1}$
\citep{2011PASJ...63..489S}.

\subsection{Global variability properties}
\label{Resu1}

W Comae was observed between 2008 February 28 and 2013 May 17. 
During this period, 32 observing runs of seven nights per run
were carried out, around the new moon phase; in total, we collected 141 data
points. The observational results are presented in Table~\ref{tbl-1} where
Column 1 is the observation cycle (see explanation in next paragraph); Columns
2 and 3 give the Gregorian and Julian Date of the observation,
respectively; Columns 4 and 5 give the polarization degree and its error,
respectively; Columns 6 and 7 give the orientation of the electric vector
position angle (EVPA) and its error, respectively; Columns 8 and 9 give the
$R$--band magnitude and its error, respectively, and; Columns 10 and 11 give
the $R$--band flux and its error, respectively.

\begin{figure}[ht]
\epsscale{0.80}
\plotone{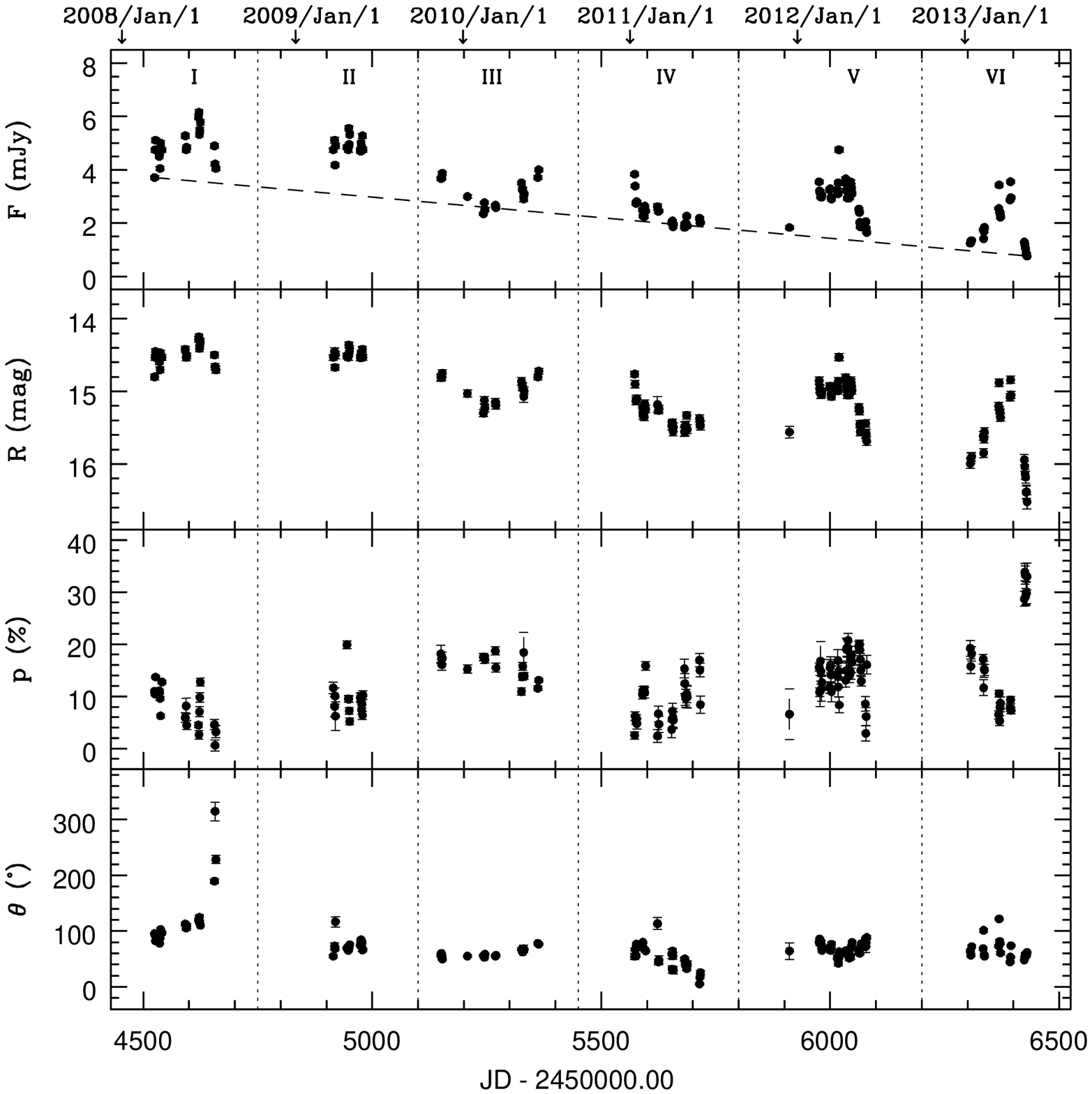}
\caption{ Photopolarimetric light curves of W Comae from 2008~February~28 to
  2013~May~17 : (from top to bottom) $R$--band flux, F(mJy); magnitude, $R$
  (mag); percentage of linear polarization in the $R$--band, $p(\%)$; and
  orientation of the EVPA, $\theta$($^{\circ}$). Vertical dashed lines
  separate the monitoring period into Cycles I to VI. Associated errors are
  presented in Table~\ref{tbl-1}. $R$--band magnitudes and fluxes have been
  corrected for the host galaxy contribution (see text for explanation).
 The $R$-band light curve (top panel) shows a slow decreasing of the mean level flux (dashed line) with a fall of $\sim$3 mJy in 5.2 yr.
 \label{fig1}}
\end{figure}

Figure~\ref{fig1} shows the $R$--band flux and magnitude light curve, the
percentage of linear polarization, $p$, and EVPA, $\theta$, obtained in a 
period of $\sim$5.2~yr.  For clarity in the discussion,
the entire period of observations has been divided into six main cycles: Cycle
I from 2008 February 28 to 2008 July 11; Cycle II
from 2009 March 24 to 2009 May 28; Cycle III from
2009 November 14 to 2010 June 16; Cycle IV from 2011
January 11 to 2011 June 4; Cycle V from 2011 December 15
to 2012 June 1; and Cycle VI from 2013 January 13 to
2013 May 17. These cycles are marked with dashed
vertical lines in Figure~\ref{fig1}, and they will be discussed in more detail
in the next paragraphs.

The statistical data analysis of the four main observational parameters
($R$--band magnitude and flux, degree of linear polarization and EVPA) was
done following \citet{2013ApJS..206...11S}. The analysis provides the average
value, the maximum and minimum observed values, and the maximum variation of
the parameters. To find out the variability in flux, degree of linear
polarization, and polarization position angle, a $\chi^2$-test was carried
out.

The amplitude of the variations $Y(\%)$ was estimated using flux densities
instead of magnitude differences following
\citet{1996A&A...305...42H},
\begin{equation}
  Y(\%) = \frac{100}{\cal h S i}\sqrt{(S_{\rm max}-S_{\rm min})^2-2\sigma^2_c} \;\; ,
\end{equation}
where $S_{\rm max}$ and $S_{\rm min}$ are the maximum and minimum values of the flux
density, respectively. ${\cal hSi}$ is the mean value, and $\sigma^2_c =
\sigma^2_{\rm max}+\sigma^2_{\rm min}$. The variability is described by the
fluctuation index $\mu$ defined by
\begin{equation}
\mu = 100\frac{\sigma_S}{\cal hSi}\% \; ,
\end{equation}
and the fractional variability index of the source $\cal F$ obtained from the
individual nights:
\begin{equation}
{\cal F} = \frac{S_{\rm max}-S_{\rm min}}{S_{\rm max}+S_{\rm min}} \; .
\end{equation}
We have estimated the minimum flux variability timescale using the definition
proposed by \citet{1974ApJ...193...43B}:
\begin{equation}
 \tau=dt/\ln(F_1/F_2) \;\; ,
\end{equation}
where $dt$ is the time interval between flux measurements $F_1$ and $F_2$,
with $F_1>F_2$. We have calculated all possible timescales $\tau_{ij}$ for any
pair of observations for which $\mid F_i-F_j\mid >\sigma_{F_i}+\sigma_{F_j}$.
The minimum timescale is obtained when:
\begin{equation}
 \tau_{\rm var}=\mbox{min}\{\tau_{ij,\nu}\} \; , 
\end{equation}
where $i=1,...,N-1; j=i+1,...,N,$ and $N$ is the number of observations. The
uncertainties associated to $\tau_{\nu}$ were obtained through the errors in
the flux measurements.

Table~\ref{tbl-2} shows the results obtained from the statistical analysis:
Column 1 gives the corresponding cycle; Column 2 the variable parameters;
Columns from 3 to 10 present, for each of the four parameters, its average, the
maximum and minimum observed value, the maximum variation $\Delta_{\rm max}$,
the variability amplitude $Y(\%)$, the variability index $\mu$(\%), the
variability fraction $\cal F$, and the statistic $\chi^{2}$, respectively.  We
have estimated the minimum flux variability timescale of $\tau_{\rm var}$ =
3.3$\pm$0.3 d.

\subsection{Photometric variability} 

Considering the entire data set, a brightness maximum of $R=$14.25 mag
was observed in 2008 Jun 4 and a brightness minimum of $R=$16.52 mag in
2013 May 17. A variation of $\Delta\,m_{R}=$2.27 mag (5.40 mJy) in $\Delta
t=$1905~d ($\sim$5.2 yr) is found (see Table~\ref{tbl-2}). During our
monitoring period, the source showed a maximum brightness variation in
timescales from months to years. There can be noticed a tendency of a slow
decreasing brightness after each flare episode, which is shown in
Figure~\ref{fig1}. In this figure a fall of $\sim$ 3 mJy in
$\sim$5.2 yr, superimposed on rapid brightness variations with timescales of
months and days, can be seen. The time between peak brightness maxima is $\sim$ 0.9-1.0 yr.

The most important photometric results are found in Cycles I, V and VI (see Table~\ref{tbl-2}).
In Cycle I W Comae shows a maximum flux of 6.16$\pm$0.10~mJy
in 2008~June~4. This flare lasted $\sim$2 months. A minimum flux of 3.71$\pm$0.07~mJy 
is observed in 2008 February 28. The flux changed 2.45~mJy in 97~days. We want to point out 
here that all photometric $R$-band data collected in 2008 are already published in \citet{2009ApJ...707..612A}.
In Cycle V the source presented the maximum flux variability of 3.10~mJy (1.15~mag) 
in 60 days.  In 2012 March 30 the source brightened 1.65 mJy in 3 days.
Finally, in Cycle VI the source presented a change in flux of 2.80 mJy 
in a period of 36 days.  It is important to note that the observed flux variations in this cycle 
correspond to a long-term flare ($\sim$~4 months). 
In this long-term flare there are two superimposed short-term flares (3.43 mJy and 3.55 mJy)
with a duration of three days each.

\subsection{Polarimetric variability}
\label{Polvar}

\subsubsection{Polarization degree variability}

\begin{figure}[ht]
\epsscale{0.80}
\plotone{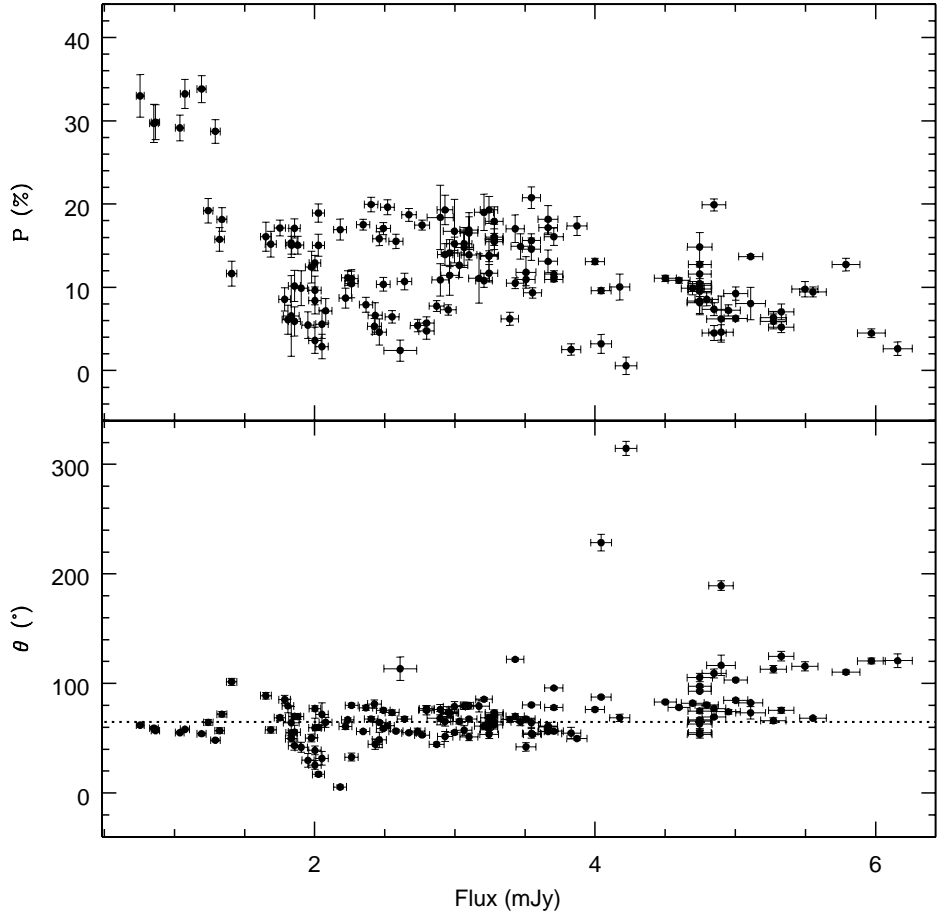}
\caption{Correlations between photopolarimetric observations of W Comae for
  all data. Top panel: correlation between the $R$--band flux and the
  polarization degree. Bottom panel: correlation between the $R$--band flux
  and EVPA. The dotted line at the bottom panel shows the preferred EVPA
    of 65\degr .\label{fig2}}
\end{figure}

Figure~\ref{fig2} shows the correlations between the flux and the polarization
degree (top panel), and the flux and the EVPA (bottom panel), for all
cycles.  To establish a possible correlation between the polarization degree
and the $R$--band flux, a Pearson's correlation coefficient was calculated
($r_{F-p}$). This coefficient was tested through the Student's $t$-test. Using
all data,  we found that there is no correlation between the
$R$--band flux and the polarization degree (see top panel of
Figure~\ref{fig2}). However, the degree of polarization shows a slight
tendency to increase as the brightness decreases.
In Table~\ref{tbl-3}  the results of the statistical analysis for the correlations 
between flux and polarization (both on the percent of polarization and EVPA)
are presented.

We did not find any correlation between the $R$--band brightness and the polarization degree, except for the
Cycles II, III, and VI where a moderate anticorrelation exists. 
In Cycle II, the Pearson's correlation coefficient is $r_{F-p}=-0.88\pm0.24$ 
during the fall of the flare. In Cycle
III its value is $r_{F-p}=-0.82\pm0.09$ during the rise of the flare.  In
cycle VI, $r_{F-p}=-0.89\pm0.04$ (taking into account the rise and fall of the
flare). This result points out that both the flux and the polarization degree show
a tendency to be anti correlated in periods of time $\sim$~weeks-months. On
the other hand, a positive correlation of $r_{F-p}=0.93\pm0.11$ was found
during Cycle I (2008 June 3-7 flare).  In general, the polarization degree
showed a random variability behavior, with a maximum and a minimum of
(33.8$\pm$1.6)\% (2013 May 12) and (0.6$\pm$1.0)\% (2008 July 9), respectively. 
The maximum variability observed was $\Delta
P$=33.2\%, in $\Delta t = $1768~days ($\sim4.8$~yr). It is interesting to
notice that the maximum value of the polarization degree occurred in Cycle VI,
when the brightness was at its minimum.

The maximum and minimum polarization degrees for each cycle are shown in the
Table~\ref{tbl-2}. In Cycle I, the maximum variability observed of the
polarization degree is $\Delta P=13.1\%$ in $\Delta\,t=130$ days; in Cycle II,
$\Delta P= 14.7\%$ in $\Delta t$ = 6 days; in Cycle III, $\Delta P= 7.8\%$ in
$\Delta t$ = 57 days; in Cycle~IV, $\Delta P= 14.5\%$ in $\Delta t$=92 days;
in Cycle V, $\Delta P= 17.8\%$ in $\Delta t$=39 days; and in Cycle VI, $\Delta
P=28.5\%$ in $\Delta t$=55~days.

\subsubsection{Position angle variability} 

In general, our data do not show a clear correlation between the polarization
angle and the $R$--band flux (see bottom panel of Figure~\ref{fig2}). Rather,
 after the large rotation observed during Cycle I,
the polarization angle presents a preferential position of $\sim$65$\degr$
(see Section 3) with maximum variations of $\Delta \theta\sim 54\degr$ (see bottom panel of Figure~\ref{fig1}).

\begin{figure}
\epsscale{0.80}
\plotone{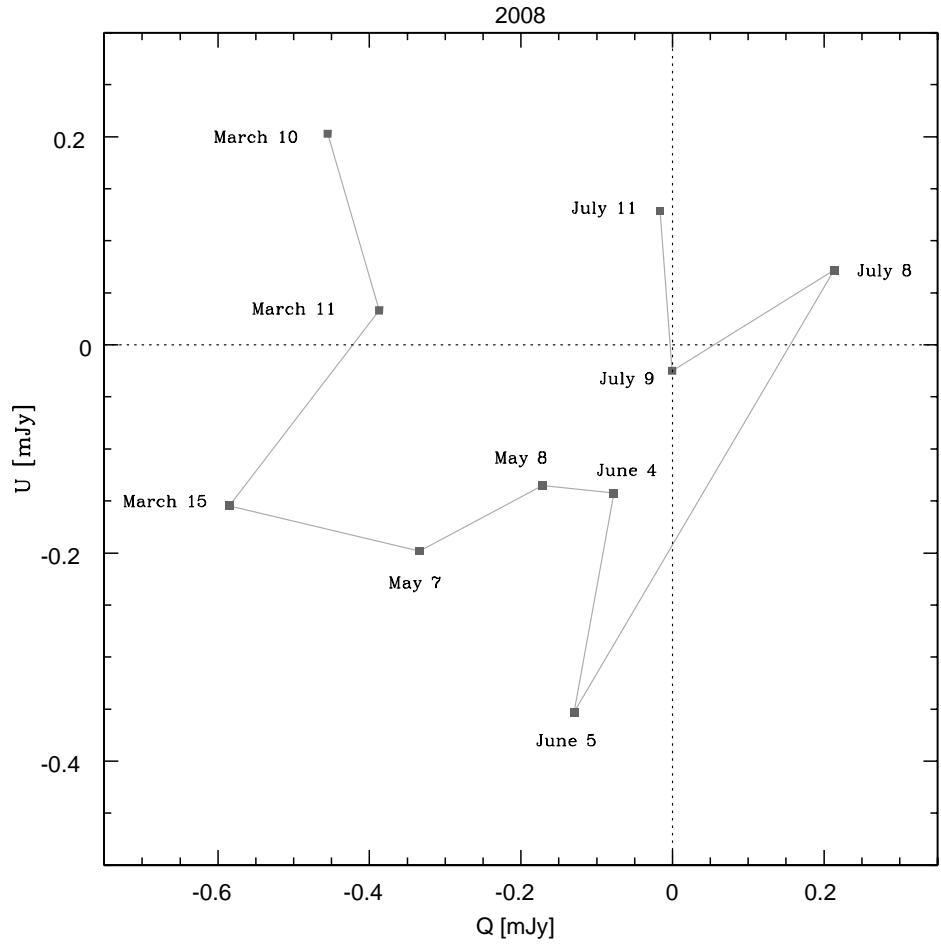}
\caption{Stokes plane showing the rotation of position angle (EVPA) of the polarization during the flare of 2008 June. 
  \label{fig3}}
\end{figure}

In Cycle I a gradual rotation of the EVPA of 78\degr \,(2008 March 10) to
315\degr (2008 July 9) is observed. This corresponds to a total rotation of
$\sim$237\degr\, in a period of 121 days (giving an average rate of rotation of $\sim$ 2\degr\, per day). 
Figure~\ref{fig3} shows this rotation in the Stokes plane. For more clarity only the more representative points are shown.

In Cycles II to VI, our data show that EVPA have the preferential value mentioned above
with mean variations rate $\sim$1.2\degr per day. In cycle IV, the EVPA reach a 
maximum value of 114\degr\, while the polarization degree 
is at its minimum value of 2.4\%. And the other way around, when the EVPA shows its 
minimum value of 6\degr, the polarization degree 
shows its maximum value of 16.9\%.

\section{POLARIMETRIC ANALYSIS}
\label{Polana}

From our observations we have found that W Comae shows, in general, a random
polarimetric behavior. This has been explained as due to the presence of
one or more variable polarization components overlaid on a stable one.
To identify the presence of a stable polarized component, we have used the method suggested by
 \citet{1985ApJ...290..627J}. In this work, the authors proposed that if the
 observed average values ($\cal h$ $Q$ $\cal i$, $\cal h$ $U$ $\cal i$) in
 the absolute Stokes parameters plane $Q$-$U$ deviate significantly from the
 origin, then a stable polarization component is present. From our data, the derived average values of the absolute Stokes parameters
are $\langle Q \rangle$ = -0.22$\pm$0.02 mJy and $\langle U \rangle$=
0.21$\pm$0.03 mJy. These average values correspond to a stable component with
constant polarization degree $P_c= 10.7\%\pm0.8\%$ and polarization angle
$\Theta_c=$65\degr$\pm$2$\degr$. The constant polarization degree has a
dispersion $\sigma_{P_{c}}=$ 6.4\%.

\begin{figure}
\epsscale{1.0}
\plottwo{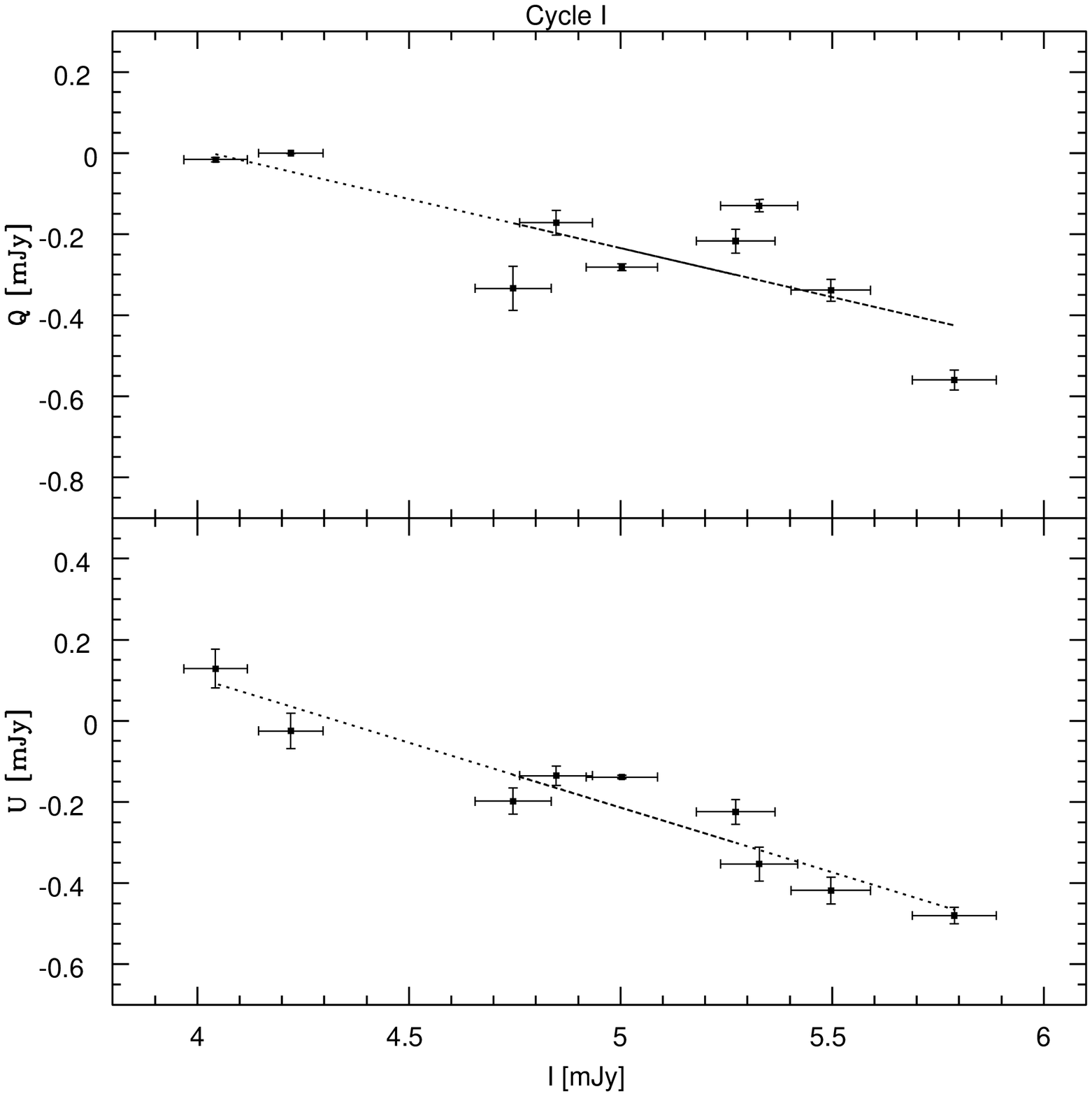}{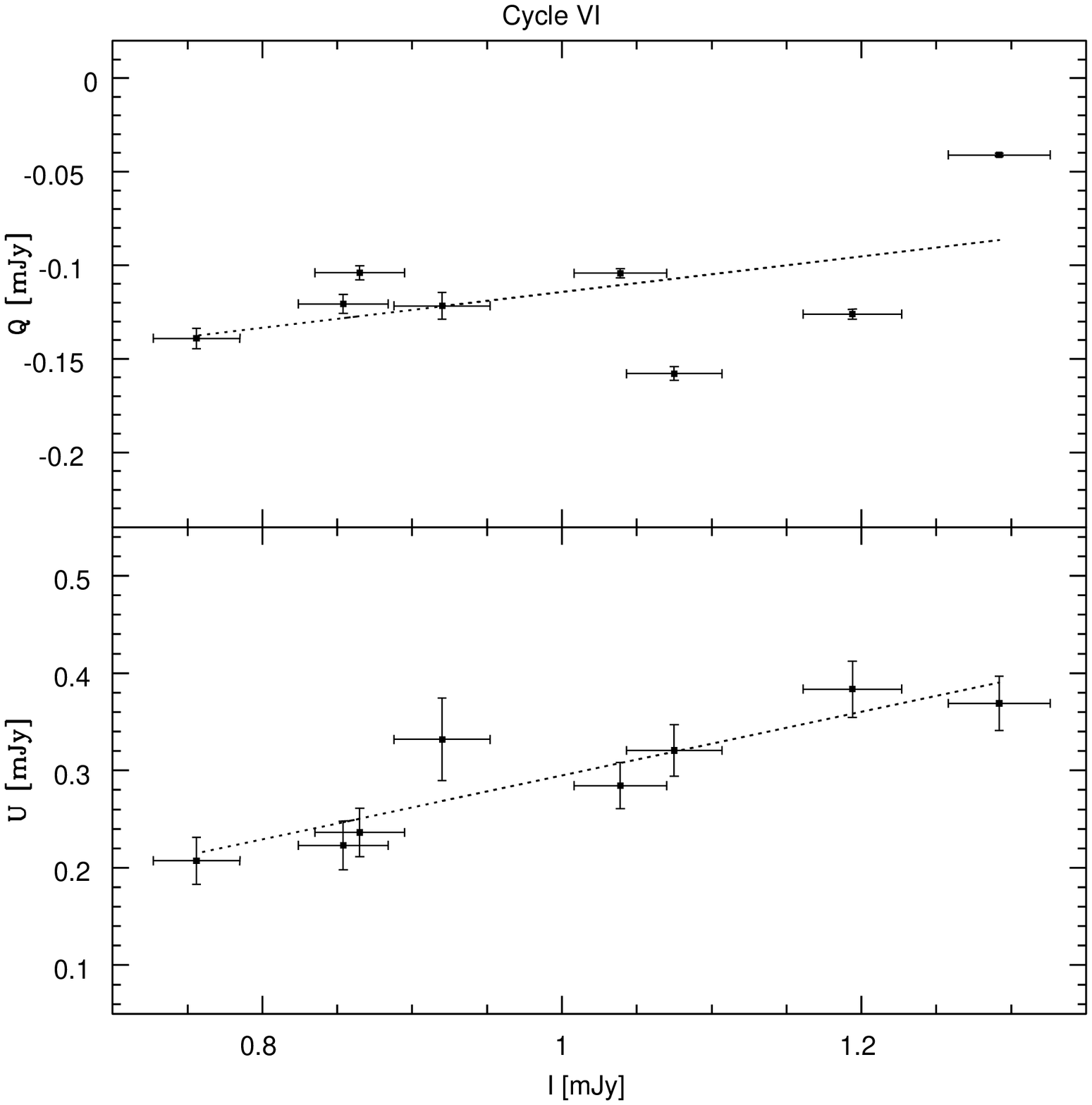}
\caption {Left: linear correlation between the Stokes parameter $Q$ vs $I$ (top panel), and $U$ vs $I$ (bottom panel) for Cycle I. The correlation coefficients for this cycle are $r_{QI}$=0.80 and $r_{UI}$=0.96, and the slopes $m_{QI} =-0.24\pm 0.07$ and $m_{UI}=-0.32\pm0.04$, respectively. Right: linear correlation between the Stokes parameter $Q$ vs $I$ (top panel), and $U$ vs $I$ (bottom panel) for Cycle VI. The correlation coefficients for this cycle are $r_{QI}$=0.51 and $r_{UI}$=0.89, and the slopes $m_{QI} = 0.10\pm 0.07$ and $m_{UI}= 0.33\pm0.07$, respectively.
  \label{fig4}}
\end{figure}

To estimate the polarization variable component parameters, we looked for a
possible linear relation between $Q$ versus $I$ and $U$ versus $I$ for the six
relevant cycles \citep[see, ][]{2008ApJ...672...40H}. For Cycle IV no linear
correlation between these parameters was found; rather, they appear to be
randomly related.  In contrast, for Cycles I, II, III, V, and VI, our data
show a linear tendency between these parameters. We made a least square fit to
the data in order to find the slopes and the linear correlation coefficients
$r_{QI}$ and $r_{UI}$.  Figure~\ref{fig4} shows this linear correlations
between Stokes parameters for cycles I and VI.  The correlation coefficients
for these parameters are given in Table~\ref {tbl-4} where Columns 2 to 7 give
the parameters $q_{var}$, $r_{QI}$, $u_{var}$, $r_{UI}$, $p_{\rm var}$ and
$\theta_{\rm var}$, respectively. The maximum polarization degree found for the
variable component is {\bf $p_{\rm var}^{\rm max}\,=\,(40.1\pm5.1$)\%}, with a polarization angle
$\theta_{\rm var}\,=\,116\degr\pm7\degr$, corresponding to Cycle I.

\subsection{The two-component Model}
\label{2comp}

From the above results we infer the presence of a stable component that we
assume associated with the relativistic jet, and also a variable component
that can be related to the propagation of a shock. 
Therefore, the observed polarization would be the result of the overlap of these two optically thin synchrotron components.

Assuming that there are two polarimetic components in W Comae, we have used equations (1) and (2) 
given in \citet{1984MNRAS.211..497H} and derive the following equations for the parameters associated to the polarized variable component:

\begin{equation}\label{pvar}
  p^2_{\rm var}=\frac{p_{\rm cons}^2+p^2( 1+ I_{v/c})^2 -2pp_{\rm cons}(1+I_{v/c}) \cos2(\theta_{\rm cons}-\theta)} { I_{v/c}^2}
  \; \; ,
\end{equation}
and
\begin{equation}
\tan2\theta_{\rm var}=\frac{ p(1+ I_{v/c}) \sin2\theta - p_{\rm cons}\sin2\theta_{\rm cons}} {p(1+ I_{v/c}) \cos2\theta - p_{\rm cons}\cos2\theta_{\rm cons}}
\;\; .
\end{equation}

where $I_{v/c}$ is the flux ratio between the variable to the constant component, and $p$ and $\theta$ are observed polarimetric parameters.
This system of equations has five free parameters: $p_{\rm cons}$, $\theta_{\rm cons}$, $p$, $\theta$ and $I_{v/c}$.  The system can be resolve 
if $p_{\rm cons}$ and $\theta_{\rm cons}$ correspond to $P_{\rm cons}$ and $\Theta_{c}$ previously obtained in section~\ref{Polana}. 

To obtain $I_{v/c}$, we maximize equation (~\ref{pvar}) with respect to $\theta$. From our observacions, $p_{\rm var}$ reaches maximum values when $p \geq p_{\rm cons}$ 
and $\pi/2 \leq 2\,(\theta_{\rm cons} - \theta) \leq \pi$. From the analysis of the Stoke's parameters in Cycle I, we find maxima values for $p_{var}=$40\% (see Table~\ref{tbl-4}). 
This maximum occurs in 2008 June 7, with $p=$12.7\% and $\theta=110\degr$, just a day before the huge gamma-ray flare. With these values we estimate
$I_{v/c}=0.57\pm0.07$.  Applying the same procedure in Cycle VI, where the blazar presents a minimum activity state, we find that $I_{v/c}=3.98\pm0.32$.

\begin{figure}
\epsscale{1.0}
\plottwo{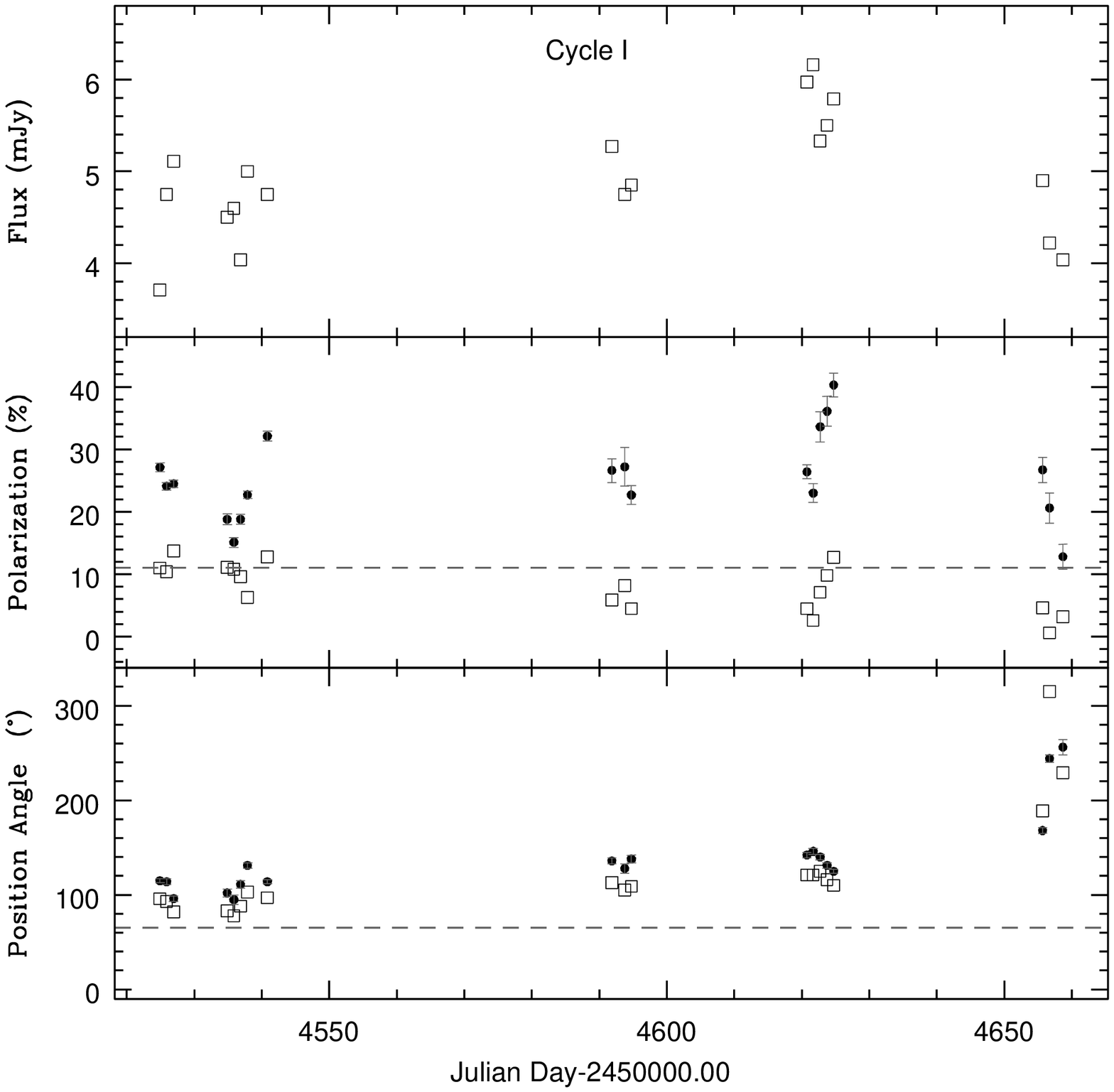}{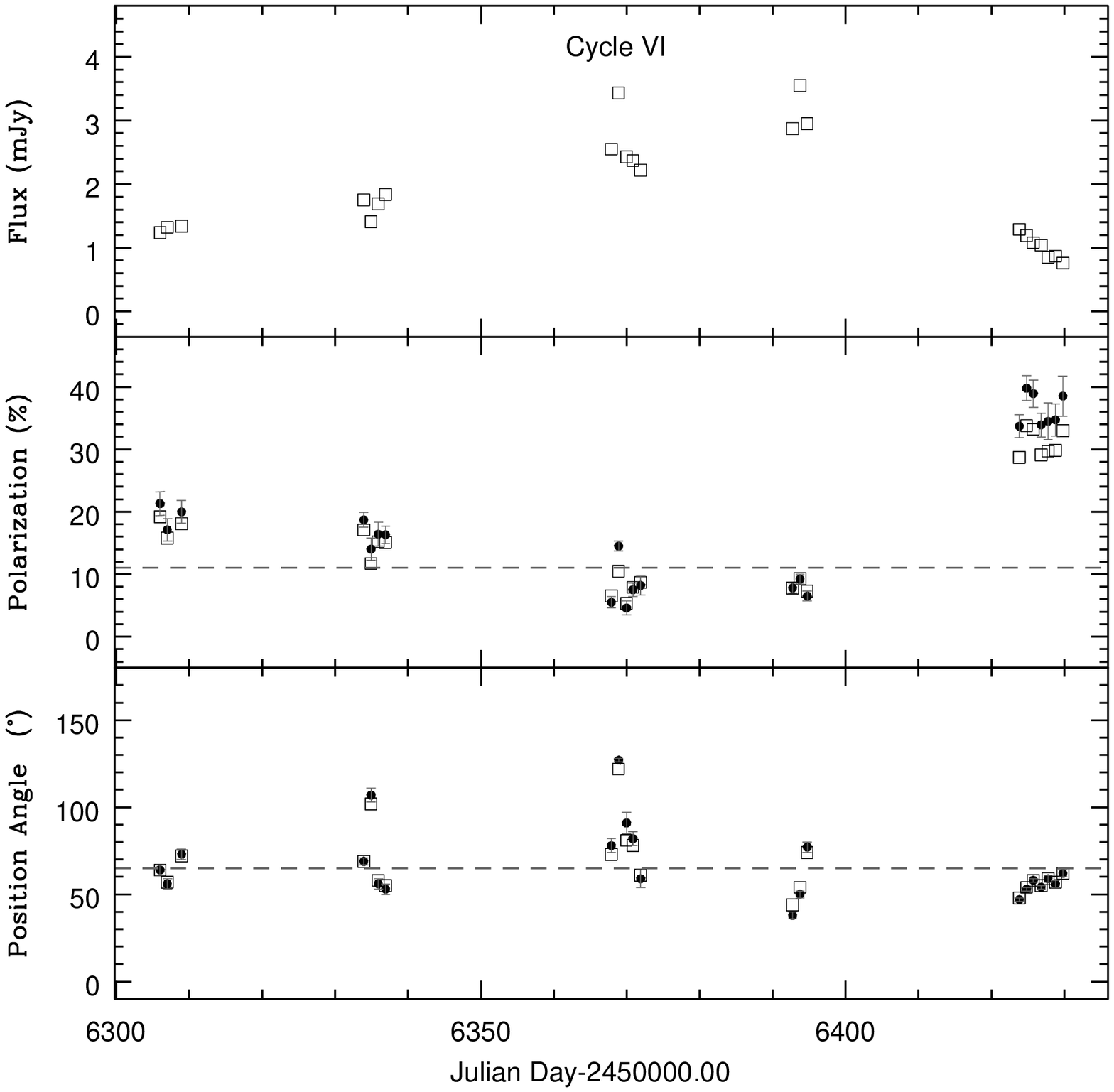}
\caption{Left and right: empty squares show the observed polarization due to the contribution of the two polarized components, one with constant polarization (dashed line) and another with variable polarization (solid dots).
Upper panels show the variations of observed flux. Middle panels, the variations of the the polarized degree. Lower panels, the variations of the  EVPA. 
Left side: In Cycle I the variable polarized component follows the variations of the observed flux, while the observed polarization is weak correlated with it.
Right side: In Cycle VI  the variations displayed by the the observed polarized degree are followed by the variable polarized component. For more details see Section~\ref{2comp}.
\label{fig5}}
\end{figure}

The values of $p_{\rm var}$ and $\theta_{var}$ are shown in Figure~\ref{fig5}, where the observed polarization $p$ is the combination of the two polarization 
components (stable plus variable). For Cycle I (high activity state) it can be seen that the variable polarization component $p_{\rm var}$ shows a similar variability 
behavior as the observed flux in the R-band. We previously assumed that this variable polarization component is associated with the propagation of shocks along the jet. 
In the same figure, we show the results for Cycle VI (low activity state) where the observed polarization $p$ and the variable polarization component 
$p_{\rm var}$ show a similar variability behavior. It is interesting to note that $\theta_{\rm var}$  follows the observed EVPA variations in both cycles.

\section{DISCUSSION}
\label{Dis}

In this work, we have inferred the presence of two components to explain
the optical polarization variability.  The minimum variability scale of
$\sim$3 days was found and it is superimposed on a longer-term flare that
lasts $\sim$3 months (these long-term flares appeared separated by $\sim$0.9
yr).  The variability timescales found in this work are in agreement with
previous studies \citep{1998A&AS..130..109T,1999A&A...342L..49M}.
 
In 2008 June 8 a strong outburst of very high energy gamma-ray emission above 200 GeV, was detected with
{\it VERITAS} in W Comae with a significance of 10.3
\citep{2009ApJ...707..612A}.  Data from our monitoring for 2008 June 4-7 show
an increase in the $R$-band flux. 
Unfortunately, due to bad weather we could not obtain data for June 8, when the maximum
brightness was observed in the $\gamma$-rays. However, our data show a gradual
increase in the value of the EVPA from 78\degr  to  315\degr~(2008 March 10--2008 July 9)
and a large rotation of $\sim$237\degr during cycle I, coinciding with the
2008 flare.

The large rotation of EVPA can be interpreted as due to an asymetric
distribution of the magnetic field with respect to the jet axis. 
\citet{2001A&A...374..435M} show that the jet has a spiral structure at 1.6 and 5 GHz. 
On the other hand, \citet{1994ApJ...435..140G} suggest that the polarization degree and the
different values of the EVPA from their VLBI images can be due to shocks propagating along
a curved jet, producing an ordered magnetic field with helical structure. These studies
found that the jet of W Comae has a projected position angle of $\sim110^{\circ}$ at 1.6 GHz and 5 GHz.
Therefore, the rotation can be produced by a swing of the jet
along the visual line of sight, or a curved trajectory of the
dissipation/emission pattern. In agreement with \citet{2010Natur.463..919A},
the second possibility may be due to the propagation of a knot emission which
follows a helical path in a magnetically-dominated jet or can be due to an
entire bending of the jet.

The direct association found between the $\gamma$-ray flare in 2008 and the
gradual change in the EVPA suggests that the $\gamma$-ray and optical emission
regions are co-spatial.  This implies a highly ordered magnetic field in
regions where the $\gamma$-rays emission is produced, therefore this
strong flare could have been produced by a strong shock.  Taking into
account the properties mentioned above, we assumed that the strong flare
observed in 2008 in optical and in $\gamma$-rays is a combination of two
factors. On one hand, if a curved structure of the jet is assumed,  the jet
direction will be oriented towards the observer with the minimum viewing angle.
On the other hand, a strong shock occurred at the same time. We will discuss
this hypothesis in the following section.

\subsection{Alignment of magnetic field by the Shock}

From our results, the moderate anti--correlation found in some flares between
the flux and the polarization degree indicates that the magnetic field tends
to be aligned with the jet.  This result is in agreement with
\citet{1996MNRAS.283..759G}. However, during the 2008 June major flare,
lasting in $\gamma$-rays only three days, the flux correlates with the
polarization degree thus suggesting that this event was originated by a
transversal shock. 

Thus in the observer's reference frame, the flux of the shocked
region is amplified as: 
\begin{equation}
F=F_0\nu^{-\alpha}\delta^{(3+\alpha)} \;\;,
\label{F0F}
\end{equation}
\citep[see][]{2009herb.book.....D}
where $\delta=[\Gamma_{j}(1-\beta\cos\Phi]^{-1}$ is the jet's Doppler factor,
$\beta = (1-\Gamma_{j}^{-2})^{1/2}$ its global velocity in units of speed of
light, $\Phi$ is the viewing angle, and $\alpha$ is the spectral index
in the optical bands.

The observed degree of polarization $p$ depends on the rest--frame angle between the line of sight and the compression axis $\Psi$, the
spectral index $\alpha$, and the shock compression factor $\eta$, which is the ratio of densities of a plasma of 
 relativistic electrons from the shocked to the unshocked region $\eta =
\eta_{\rm shock}/\eta_{\rm unshock}$  \citep{1991bja..book....1H}:
\begin{equation}
  p\approx\frac{\alpha+1}{\alpha+5/3}\frac{(1-\eta^{-2})\sin^2\Psi}{2-(1-\eta^{-2})\sin^2\Psi} \;\; ,
\label{F1F}
\end{equation}
and
\begin{equation}
  \Psi=\tan^{-1}\left\{\frac{\sin\Phi}{\Gamma_{j}(\cos\Phi-\sqrt{1-\Gamma_{j}^{-2}})}\right\}
  \;\; .
\label{F2F}
\end{equation}
Following \citet{2009ApJ...707..612A}, we assumed a bulk Lorentz factor
$\Gamma_j$ = 20 for W Comae. We also used the value of $\alpha$ = 0.87, given by \citet{1998A&AS..130..109T}.

From equation ~(\ref{F0F}), we can estimate the Doppler factor as a function of time. The value of $F_0$ is determined by $F_0 = F_{\rm max} \nu^{\alpha} /\delta_D^{(3+ \alpha)}$, where $F_{\rm max}$ is the maximum observed flux and  $\delta_D$ is obtained from $\Phi_0$, which is calculated from equations~(\ref{F1F}) and~(\ref{F2F}) for $p\approx p_{\rm var}^{\rm max}$ this being the maximum value of the polarization degree of the variable component (see Table~\ref{tbl-4}).  From \citet{1991bja..book....1H}, for $\Psi =\pi/2$, $\eta$=2.2 which is the minimum compression that produces a degree of linear polarization as high as 45$\%$.  This yields to $\Psi_0\approx 70\degr$, $\Phi_0\approx 2.0\degr$, and $\delta_D\approx$26.7 at the maximum polarization of the variable component. Using equations (\ref{F0F}), (\ref{F1F}), and (\ref{F2F}) the physical parameters $\delta, \Phi, \Psi$, and $\eta$ as a function of time were estimated.  

\begin{figure}
\epsscale{0.80}
\plotone{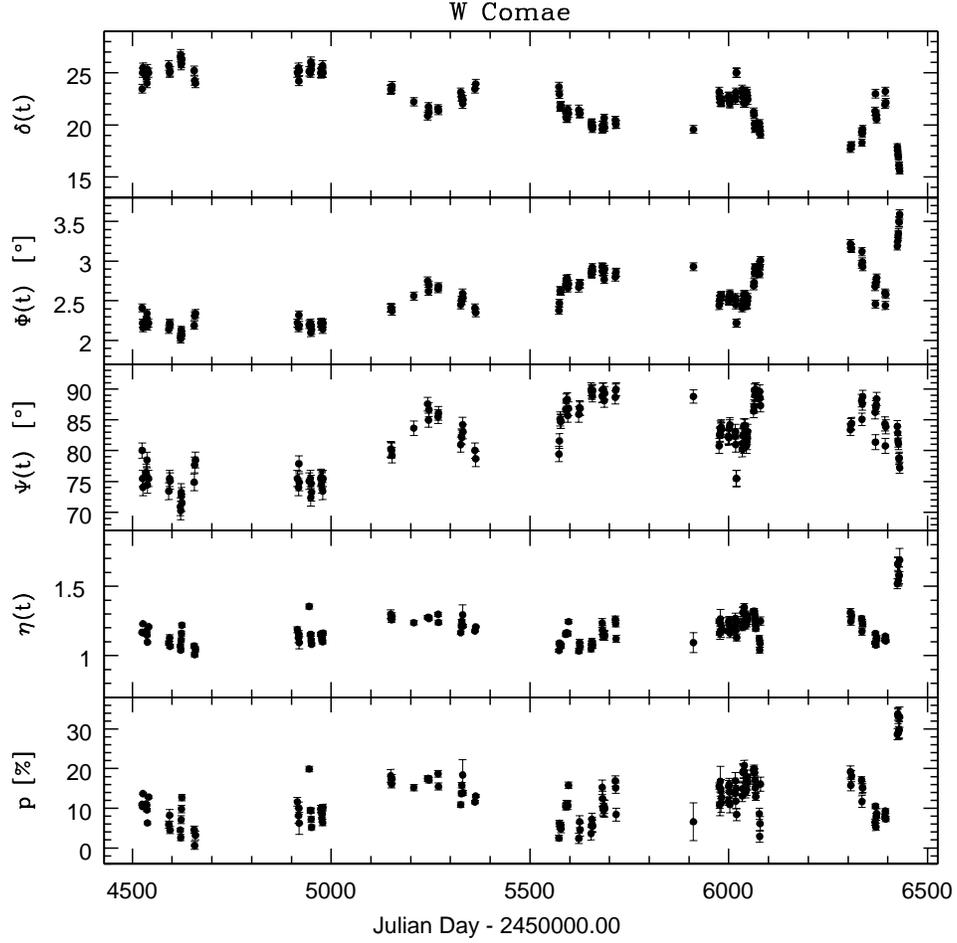}
\caption{Temporal variability of some physical parameters related to the
  relativistic jet of W Comae: (from top to bottom) Doppler factor,
  $\delta(t)$; viewing angle of the jet, $\Phi(t)$; rest-frame viewing angle of the shock, $\Psi(t)$; compression factor of the shocked to the unshocked plasma, $\eta(t)$. These parameters were estimated using the maximum flux value observed during Cycle I.  Finally, the lower panel shows the degree of linear polarization, $p$[\%], in the $R$-band.}
  \label{fig7}
\end{figure}

In Figure~\ref{fig7}, it can be seen that the source shows its maximum brightness (14.25 mag, 2008 June 4 or JD 2454621), and the Doppler factor reaches 26.7, while during the minimum (16.5 mag, 2013 May 13 or JD 2456429) it is 15.6. This corresponds to a maximum variation of $\Delta\delta~\sim$11. The viewing angle of the jet $\Phi$, shows a minimum value of 2\degr.0 and a maximum value of 3\degr.6, i.e., $\Delta\Phi~\sim$1.6.  These small variations of the Doppler factor can produce large flux variations while $\Gamma_j$ remains constant. In the state of maximum brightness, the viewing angle of the shock $\Psi\sim70\degr$ undergoes its maximum aberration due to relativistic effects.

A maximum compression of the plasma of $\eta$=1.69 is found, when the
polarization degree observed reaches its maximum value of $33\%$ 
(2013 May 17 or JD 2456429). The minimum compression factor $\eta_{\rm min}$=1.01 is 
obtained when the polarization degree had a minimum value of 0.6$\%$ 
(2008 July 9 or JD 245 4656). These, small changes in the compression factor 
($\Delta\eta \approx 0.68$) can produce large changes in the polarization

The Doppler factor $\delta_D$ is obtained when the $R$-band flux is at its
maximum value due to the presence of the shock. Then, the change in the
magnetic field intensity due to the shock is estimated assuming that the
minimum variability timescale is related to the shock-front thickness. This
scale is estimated considering the lifetime of the synchrotron electrons
\citep[see, e.g.][]{2008ApJ...672...40H}. The lifetime of the synchrotron
electrons for a given frequency $\nu$ in GHz is
\begin{equation}
  t_{\rm loss}=4.75\times10^2\left(\frac{1+z}{\delta_D\, \nu_{\rm GHz}\,B^3}\right)^{1/2} \,\rm days \;.
\label{F3F}
\end{equation}
where $B$ is the magnetic field in Gauss. Since $t_{\rm loss} \approx t_{\rm
  var}$, for $\delta_D$ = 26.7 and $t_{\rm var}=3.3\pm0.3$ days, equation
(\ref{F3F}) yields an estimate of the magnetic field intensity,
$B=$~0.12$\pm$0.01~G, and an upper limit for the emission region size of
$r_{b}\leq ct_{\rm var} \delta_D/(1+z)$ = (2.1$\pm$0.2)$\times10^{17}$ cm.

Finally, in cycles where no correlation was found between the flux and the
polarization degree, the flares could be possibly due to an oblique shock to
the jet's direction, or due to changes in the Doppler factor, related to
changes in the viewing angle of the jet. Therefore, three scenarios are
proposed to explain the flares observed at different timescales: 1) a shock
transverse to the jet axis, ordering the magnetic field parallel to the
shock's plane; 2) an oblique shock with respect to the jet axis produced in an
initially disordered magnetic field, produces a final magnetic field with a
component almost parallel to the jet axis; 3) variations of the Doppler
factor due to changes in the jet axis orientation with respect to the
observer's line of sight.

From the polarimetric analysis, we found that the behavior of the polarized
variable flux could be due to the superposition of two optically--thin
synchrotron components. One stable with $\theta_{\rm cons}\approx$ 65\degr,
$p_{\rm cons}\approx$11\% and the other variable ~(see Figure~\ref{fig5}).  Assuming that the position angle of the 
radio jet is $\theta_{\rm jet}\approx$110\degr, we propose that the transversal 
shocks to the jet axis could be related to the variable component and the oblique 
shocks to the stable component.  Nevertheless, both variable and stable 
components can be affected by variations of the Doppler factor.

\section{CONCLUSIONS}
\label{Conc}
From the photopolarimetric observations of W Comae we found that the source displayed activity during the monitored period. We clearly detect four flares, estimating that the object has a minimum variability scale of 3.3 days and a maximum variability in brightness of 2.27 mag. The maximum degree of linear polarization reached by W Comae during the campaign was 33.8\%.

An important observational result is the large rotation of EVPA of  $\Delta \theta \sim 237^{\circ}$, associated to the optical flare and coincident with the 
major $\gamma$-ray flare observed in 2008 June.  Subsequently, the polarization angle tends to a preferential orientation of $\sim65^{\circ}$.  The large rotation associated with the flare in $\gamma$-rays suggests that both optical / $\gamma$-ray emissions could be produced in the same jet's region.

From the analysis of the Stokes parameters, we infer the presence of two optically thin synchrotron components with different polarimetric characteristics: one is a variable component and the other one is stable with a constant degree of polarization of $p_{\rm cons}\approx$11\%, and a constant position angle of $\theta_{\rm cons}\approx$ 65\degr. Assuming that the 2008 June optical flare has originated in a transversal shock propagating down a twisted jet, and that the source is a spherical blob of radius $r_b$, moving with a Lorentz's factor of $\Gamma=20$, from our polarimetric data we estimated a Doppler factor of $\delta_D\sim 27$ when the flux was maximum, and a visual angle of the jet $\Phi\sim 2^{\circ}.0$. We also obtained a magnetic field intensity $B \sim$ 0.12 G. Finally, an upper limit for the size of the emission region of $r_b \leq 2\times10^{17}$~cm was estimated.

The variability timescales displayed by W Comae show two main characteristics: (1) There are two components in the light-curve, one contributing to the long-term brightening with timescales going from 2 to 4 months, and the other that contributes to the short timescale variations ($\sim3$ days). This result is in agreement with  \citet{2002A&A...395...11T}. (2) The Doppler factor changes ($\delta(t)\approx16-27$) could be due to changes in the viewing angle of the jet, implying flux variations lasting $\sim$0.9 yr.

Based on the the anticorrelation found between the polarization percentage and the flux, we propose that the observed long-term flux behavior can be explained with a spiral jet and a transversal shock-wave models. This  anticorrelation depends on the Doppler factor time-variations $\delta(t)$ for a range of values of the viewing angle $\theta(t)$.

From our observations, we found that the EPVA in the optical has a value $\sim$ 110\degr\, in 2008 June 7, a day before the gamma-ray flare. This value is identical to the projected angle of the radio jet found by \citet{1994ApJ...435..140G}  and \citet{2001A&A...374..435M}. Later, after the strong gamma-ray flare finished, the EVPA increases its value up to 315\degr\, in 2008 July 9, and two days after it goes down to 229 deg.  During the following cycles the EVPA shows a preferential value of 65\degr. In a future work, it would be useful to measure the direction of EVPA rotations using also radio data. This will allow us to verify whether the behavior of the EVPA in the optical bands is similar to the EVPA variations studied in the radio-bands.

On the other hand, although \citet{2008PASJ...60..145Z} predicts a flare around 2013, from our data collected in 2013 we did not detect any important outburst in W Comae. Rather, we report a continuos brightness decrease detected since the beginning of 2008, reaching a minimum value in 2013 May.  But, this could also be considered as a prelude to a major flare or a flare that could start at the end of 2013.

\acknowledgments  We thank the anonymous referee for helpful comments that
improved the presentation of this work. M.S., E.B., D.H., J.I.C., and R.M. acknowledge financial
support from UNAM--DGAPA--PAPIIT through grant IN116211 
and E.B through grant IN111514. We thank the OAN-SPM staff for the support
given to this project.  This research has
made use of the SAO/NASAâ Astrophysics Data System (ADS) and of the NASA/IPAC
Extragalactic Database (NED), which is operated by the Jet Propulsion
Laboratory, California Institute of Technology, under contract with the
National Aeronautics and Space Administration.

\bibliography{SorciaWcom}

\clearpage

\clearpage

\clearpage

\clearpage

\clearpage

\clearpage

\clearpage
\begin{deluxetable}{llcrrrcrcccc}
\tablewidth{0pt} 
\tablecaption{POLARIZATION AND PHOTOMETRY IN THE $R$-BAND FOR W Comae 
\label{tbl-1}}
\tablehead{ 
\colhead{Cycle}& \colhead{Date}& \colhead{JD}& \colhead{$p$}& \colhead{$\epsilon_p$}&
\colhead{$\theta$}& \colhead{$\epsilon_{\theta}$}& \colhead{$R$}&
\colhead{$\epsilon_R$}& \colhead{$\rm Flux$}& \colhead{$\epsilon_{\rm Flux}$}\\
\colhead{ }& \colhead{ }& \colhead{2,450,000.00+}& \colhead{($\%$)}& \colhead{($\%$)}&
\colhead{($\degr$)}& \colhead{($\degr$)}& \colhead{(mag)}&
\colhead{(mag)}& \colhead{(mJy)}& \colhead{(mJy)}&
}
 \startdata  
 I&2008 Feb 28 &4524.9351 &  11.0 &  0.3 &  96 & 01 & 14.80 & 0.04 &  3.71 &  0.07 \\ 
 &2008 Feb 29 &4525.9038 &  10.4 &  0.2 &  93 & 01 & 14.53 & 0.04 &  4.75 &  0.08 \\ 
 &2008 Mar 01 &4526.9429 &  13.7 &  0.2 &  82 & 01 & 14.45 & 0.04 &  5.11 &  0.09 \\ 
 &2008 Mar 09 &4534.8984 &  11.1 &  0.4 &  83 & 01 & 14.59 & 0.04 &  4.50 &  0.08 \\ 
 &2008 Mar 10 &4535.8706 &  10.8 &  0.3 &  78 & 01 & 14.57 & 0.04 &  4.60 &  0.08 \\ 
 &2008 Mar 11 &4536.8555 &   9.6 &  0.4 &  88 & 01 & 14.70 & 0.04 &  4.04 &  0.07 \\ 
 &2008 Mar 12 &4537.9004 &   6.3 &  0.3 & 103 & 02 & 14.47 & 0.04 &  5.00 &  0.08 \\ 
 &2008 Mar 15 &4540.8467 &  12.8 &  0.3 &  97 & 01 & 14.53 & 0.04 &  4.75 &  0.08 \\ 
 &2008 May 05 &4591.8618 &   5.9 &  0.9 & 113 & 03 & 14.42 & 0.04 &  5.27 &  0.09 \\ 
 &2008 May 07 &4593.7847 &   8.2 &  1.5 & 105 & 04 & 14.53 & 0.05 &  4.75 &  0.09 \\ 
  \enddata
 \tablecomments{The table is available in its entirety in a machine-readable form in the on-line journal. A portion is shown here for guidance regarding its form and content.}
\end{deluxetable}

\begin{deluxetable}{lrrrrrrrrr}
\tablewidth{0pt} 
\tablecaption{VARIABILITY PARAMETERS FOR W Comae
\label{tbl-2}}
\tablehead{
\colhead{Cycle}& \colhead{Parameter}& \colhead{Average}& \colhead{Max} &
\colhead{Min}& \colhead{$\Delta_{max}$}& \colhead{$Y(\%)$}& \colhead{$\mu(\%)$}& \colhead{$\cal F$}&
\colhead{$\chi^2$} \\
\colhead{(1)}& \colhead{(2)}& \colhead{(3)}& \colhead{(4)} &
\colhead{(5)}& \colhead{(6)}& \colhead{(7)}& \colhead{(8)}&
\colhead{(9)}& \colhead{(10)} 
}
\startdata
All  & R(mag) &   15.06 $\pm$  00.48 &  16.52 &  14.25 &   2.27 &   -    &   -    &   -    &    -    \\ 
     & F(mJy) &    3.17 $\pm$  01.26 &   6.16 &   0.76 &   5.40 &  170.1 &   39.6 &   0.78 &  62705.7 \\ 
     &  P(\%) &   12.52 $\pm$  06.35 &  33.82 &   0.59 &  33.23 &  264.6 &   50.7 &   0.97 &   4404.1 \\ 
     &$\theta(\degr)$  &   72.49 $\pm$ 33.40 & 314.59 &   5.54 & 309.05 &  425.1 &   46.1 &   0.97 &  13456.4 \\ 
\hline
I    & R(mag) &   14.50 $\pm$  00.15 &  14.80 &  14.25 &   0.55 &   -    &   -    &   -    &    -    \\ 
     & F(mJy) &    4.91 $\pm$  00.66 &   6.16 &   3.71 &   2.45 &   49.8 &   13.5 &   0.25 &   1119.3 \\ 
     &  P(\%) &    7.86 $\pm$  03.85 &  13.70 &   0.59 &  13.11 &  165.7 &   48.9 &   0.92 &    979.7 \\ 
     &$\theta(\degr)$  &  124.86 $\pm$ 58.80 & 314.59 &  77.99 & 236.60 &  188.6 &   47.1 &   0.60 &   1858.6 \\ 
\hline
II   & R(mag) &   14.50 $\pm$  00.07 &  14.67 &  14.36 &   0.31 &   -    &   -    &   -    &    -    \\ 
     & F(mJy) &    4.90 $\pm$  00.31 &   5.55 &   4.18 &   1.38 &   27.9 &    6.4 &   0.14 &    212.3 \\ 
     &  P(\%) &    9.29 $\pm$  03.32 &  19.91 &   5.20 &  14.71 &  157.6 &   35.7 &   0.59 &    291.2 \\ 
     &$\theta(\degr)$  &   74.90 $\pm$ 13.33 & 116.57 &  55.02 &  61.55 &   80.2 &   17.8 &   0.36 &    211.4 \\ 
\hline
III  & R(mag) &   14.97 $\pm$  00.18 &  15.30 &  14.72 &   0.58 &   -    &   -    &   -    &    -    \\ 
     & F(mJy) &    3.21 $\pm$  00.52 &   4.00 &   2.35 &   1.65 &   51.4 &   16.3 &   0.26 &   1262.7 \\ 
     &  P(\%) &   15.75 $\pm$  02.37 &  18.71 &  10.93 &   7.78 &   48.5 &   15.0 &   0.26 &    205.0 \\ 
     &$\theta(\degr)$  &   61.16 $\pm$  07.98 &  77.96 &  49.76 &  28.20 &   45.8 &   13.1 &   0.22 &    431.3 \\ 
\hline
IV   & R(mag) &   15.32 $\pm$  00.20 &  15.56 &  14.76 &   0.80 &   -    &   -    &   -    &    -    \\ 
     & F(mJy) &    2.34 $\pm$  00.48 &   3.83 &   1.83 &   2.00 &   85.2 &   20.5 &   0.35 &   1885.4 \\ 
     &  P(\%) &    8.65 $\pm$  04.11 &  16.93 &   2.41 &  14.52 &  165.3 &   47.5 &   0.75 &    377.0 \\ 
     &$\theta(\degr)$  &   53.21 $\pm$ 22.47 & 113.54 &   5.54 & 108.00 &  200.9 &   42.2 &   0.91 &   1970.1 \\ 
\hline
V    & R(mag) &   15.08 $\pm$  00.29 &  15.68 &  14.53 &   1.15 &   -    &   -    &   -    &    -    \\ 
     & F(mJy) &    2.96 $\pm$  00.73 &   4.75 &   1.65 &   3.09 &  104.6 &   24.6 &   0.48 &   6645.1 \\ 
     &  P(\%) &   14.31 $\pm$  04.05 &  20.75 &   2.89 &  17.86 &  123.3 &   28.3 &   0.76 &    312.8 \\ 
     &$\theta(\degr)$  &   67.91 $\pm$ 10.66 &  88.82 &  42.14 &  46.68 &   68.0 &   15.7 &   0.36 &    472.1 \\ 
\hline
VI   & R(mag) &   15.68 $\pm$  00.51 &  16.52 &  14.84 &   1.68 &   -    &   -    &   -    &    -    \\ 
     & F(mJy) &    1.82 $\pm$  00.85 &   3.55 &   0.76 &   2.80 &  153.7 &   46.7 &   0.65 &   7055.9 \\ 
     &  P(\%) &   17.86 $\pm$ 10.05 &  33.82 &   5.32 &  28.50 &  158.9 &   56.3 &   0.73 &   1136.3 \\ 
     &$\theta(\degr)$  &   66.22 $\pm$ 17.77 & 121.93 &  44.43 &  77.50 &  116.9 &   26.8 &   0.47 &   2393.4 \\ 
 \enddata
\tablecomments{There are no statistics $ Y, \mu, \cal F$ and $ \chi^2$ for the
  magnitude due to its logarithmic character.}

\end{deluxetable}

\begin{deluxetable}{lcrcc}
\tablewidth{0pt} \tablecaption{STATISTICAL ANALYSIS OF POLARIZATION AND FLUX CORRELATION
\label{tbl-3}}
\tablehead{
\colhead{Cycle}&\colhead{Relation}  &\colhead{$r$} &\colhead{t-student} &\colhead{Confidence} \\
\colhead{}&\colhead{parameters}  &\colhead{} &\colhead{} &\colhead{level} \\
\colhead{(1)}& \colhead{(2)}& \colhead{(3)}& \colhead{(4)}& \colhead{(5)} 
}
\startdata
All & F -- p            & -0.44 +/-0.02   &   5.72  &          no \\
    & F -- $\theta$     &  0.47 +/-0.01   &   6.25  &          no \\
    & p -- $\theta$     & -0.34 +/-0.02   &   4.17  &          no \\
\hline
I   & F -- p            &  0.93 +/-0.11   &   5.04  &          yes\\
    & F -- $\theta$     & -0.90 +/-0.11   &   4.15  &          yes\\
    & p -- $\theta$     & -0.92 +/-0.15   &   4.60  &          yes\\
\hline
II  & F -- p            & -0.88 +/-0.24   &   4.60  &          yes\\
    & F -- $\theta$     & -0.71 +/-0.24   &   2.47  &          no \\
    & p -- $\theta$     &  0.53 +/-0.27   &   1.54  &          no \\
\hline
III & F -- p            & -0.82 +/-0.09   &   4.24  &          yes\\
    & F -- $\theta$     &  0.81 +/-0.08   &   4.10  &          yes \\
    & p -- $\theta$     & -0.68 +/-0.04   &   2.81  &          no \\
\hline
IV  & F -- p            & -0.75 +/-0.07   &   3.18  &          no \\
    & F -- $\theta$     & -0.53 +/-0.16   &   1.75  &          no \\
    & p -- $\theta$     &  0.25 +/-0.14   &   0.75  &          no \\
\hline
V   & F -- p            &  0.16 +/-0.08   &   0.98  &          no \\
    & F -- $\theta$     & -0.39 +/-0.08   &   2.63  &          no \\
    & p -- $\theta$     & -0.22 +/-0.13   &   1.37  &          no \\
\hline
VI  & F -- p            & -0.89 +/-0.04   &   7.68  &          yes \\
    & F -- $\theta$     &  0.41 +/-0.05   &   1.81  &          no \\
    & p -- $\theta$     & -0.61 +/-0.05   &   3.07  &          no \\
\enddata
\tablecomments{Pearson's correlation coefficient, $r$, between the observed parameters: Flux (F); polarization degree (p); and polarization angle ($\theta$). In order to verify the validity of the correlation to $99\%$, we applied a t-student test.}
\end{deluxetable}

\begin{deluxetable}{lrrrrcrrrrrrrrl}
\tablewidth{0pt} \tablecaption{STOKES PARAMETERS FOR THE VARIABLE COMPONENT OF W Comae
\label{tbl-4}}
\tablehead{
\colhead{Cycle}&\colhead{$q_{var}$}&\colhead{$r_{QI}$} &
\colhead{$u_{var}$}&\colhead{$r_{UI}$}&\colhead{$p_{var}$ ($\%$)}&\colhead{$\theta_{var}$ ($\degr$)}\\
\colhead{(1)}&\colhead{(2)}&\colhead{(3)} &
\colhead{(4)}&\colhead{(5)}&\colhead{(6)}&\colhead{(7)}
}
\startdata
I    & -0.24 $\pm$ 0.07 & 0.80 & -0.32 $\pm$ 0.04 & 0.96 & 40.1 $\pm$ 5.1 & 116 $\pm$ 07\\ 
II   &  0.13 $\pm$ 0.09 & 0.48 &  0.17 $\pm$ 0.08 & 0.59 & 21.5 $\pm$ 8.3 &  26 $\pm$ 20\\ 
III  & -0.20 $\pm$ 0.04 & 0.92 & -0.20 $\pm$ 0.06 & 0.82 & 28.4 $\pm$ 5.2 & 112 $\pm$ 09\\
V    & -0.13 $\pm$ 0.04 & 0.62 &  0.18 $\pm$ 0.05 & 0.73 & 21.5 $\pm$ 4.6 &  63 $\pm$ 10\\
VI   &  0.10 $\pm$ 0.07 & 0.51 &  0.33 $\pm$ 0.07 & 0.89 & 34.1 $\pm$ 6.7 &  37 $\pm$ 10\\ 
\enddata
\tablecomments{No statistics for Cycle IV is presented  because no significant
  correlation between Q-I or  U-I relations was found. Column (6) presents the maximum values of $p_{\rm var}$ found in each cycle. Column (7) presents the values of $\theta_{\rm var}$ corresponding to the $p_{\rm var}$ maximum given in column (6). See Section~\ref{Polana}.}
\end{deluxetable}

\end{document}